\documentclass[journal=cmatex,manuscript=article,layout=onecolumn]{achemso}

\usepackage[version=4]{mhchem} 

\makeatletter
\let\l@addto@macro\relax
\makeatother
\usepackage[fontsize=10pt]{scrextend}

\usepackage{graphicx}
\usepackage{braket}
\usepackage{ulem}   
\usepackage{dsfont}
\usepackage{amsthm,amsmath,amsfonts,amssymb,verbatim,color}
\usepackage{mathtools} 
\usepackage{bbold}
\usepackage{graphicx}
\usepackage[T1]{fontenc}
\usepackage{bm}
\usepackage[colorlinks=true,citecolor=blue,linkcolor=blue,urlcolor=blue]{hyperref}

\newcommand{\bs}[1]{{\boldsymbol{#1}}} 

\def\Rt{\left(\bs{R}\left( t\right)\right)}

\def\RR{\mathbf{R}}
\def\kk{\mathbf{k}}
\def\AA{\mathbf{A}}

\SectionNumbersOn 

\title{Towards a Topological Classification of Nonadiabaticity in Chemical Reactions}

\author{Christopher Daggett}
\affiliation{Department of Chemistry, Pennsylvania State University, University Park, Pennsylvania 16802, USA}
\author{Kaijie Yang}
\affiliation{Department of Physics, Pennsylvania State University, University Park, Pennsylvania 16802, USA}
\author{Chaoxing Liu}
\affiliation{Department of Physics, Pennsylvania State University, University Park, Pennsylvania 16802, USA}
\author{Lukas Muechler}
\affiliation{Department of Chemistry, Pennsylvania State University, University Park, Pennsylvania 16802, USA}
\affiliation{Department of Physics, Pennsylvania State University, University Park, Pennsylvania 16802, USA}
\email{lfm5572@psu.edu}

\begin{document}
\begin{abstract}
The application of topology, a branch of mathematics, to the study of electronic states in crystalline materials has had a revolutionary impact on the field of condensed matter physics. 
For example, the development of topological band theory has delivered new approaches and tools to characterize the electronic structure of materials, resulting in the discovery of new phases of matter with exotic properties.
In the framework of topological band theory, the crossings between energy levels of electrons are characterized by topological invariants, which predict the presence of topological boundary states. 
Given the frequency of energy level crossings on the potential energy surface in molecules, the applicability of these concepts to molecular systems could be of great interest for our understanding of reaction dynamics. However, challenges arise due to differing quantum mechanical descriptions of solids and molecules.
Out work aims to bridge the gap between topological band theory and molecular chemistry.
We propose that the Euler Class, a topological invariant, can be used to categorize and analyse the distribution of nonadiabatic couplings on the potential energy surface. 
To exemplify this connection, we introduce a model system with two distinct regimes that are characterized by different values of the Euler Class, yet identical potential energy surfaces. 
Contrary to expectations set by the Born-Oppenheimer approximation, we propose that these two regimes don't exhibit identical dynamics, due to a qualitatively distinct distribution of nonadiabatic couplings.
\end{abstract}

\section{Introduction}
Chemical reactions are rare dynamical events in which reactants are transformed into products through the rearrangement of electrons and atoms. From the quantum chemical perspective, these processes are collectively modeled as movement across the electronic potential energy surface (PES)\cite{schlegel2003exploring}. The most widely applied theory to describe chemical reactions is Transition State Theory (TST), in which reactions may be visualized as two energetic minima representing reactants and products separated by some high energy barrier which represents a transition state\cite{laidler1983development,pechukas1981transition}. A main goal in chemistry is to understand and to predict the dynamics required to move between these minima.

For a reaction to proceed in this paradigm the reactants must obtain sufficient energy to surmount the barrier separating them from the products. Rate theories relate the energy of the transition state relative to the reactants, referred to as activation energy ($E_a$), to the speed at which a reaction can be expected to proceed\cite{hanggi1990reaction}. Though many rigorous theories are available which endeavor to quantify both the transition state energy and the rate of reaction, the important qualitative observation is that as $E_a$ increases the rate of reaction rapidly decays~\cite{Bao2017-na}. This relationship between barrier height, or transition state energy, and kinetics is the key in traditional approaches to predicting and rationalizing reaction outcomes.

Our discussion thus far implicitly relies on some notion of a “reaction path” or “reaction progress” connecting reactant to product and passing through the transition state. Concretely, these terms refer to the total geometry of reactants as they change position, first to adopt the transition state configuration and then that of the products. 
Many parametrization schemes are available to encode nuclear positions over the course of a reaction and thus express this evolution of geometry. In this framework, the concept of a pathway with lowest barrier, known as the Minimal Energy Pathway (MEP), has become a central object in the study of chemical reactions. The reasoning is that most trajectories will remain close to this pathway. 
In practice, they can be obtained from \textit{ab-initio} calculations through Nudged Elastic Band (NEB) methods or Intrinsic Reaction Coordinate (IRC) calculations, which are readily available in quantum chemistry programs~\cite{crehuet2005reaction,schlegel2011geometry,deng1994determination,jonsson1998nudged}.

The core assumptions underlying this framework are (i) the adiabatic Born-Oppenheimer approximation (BOA), which constrains the dynamics to a single PES, and (ii)  the assumption that features of the PES away from the reaction path may be neglected.
However, there exist classes of chemical reactions in which these assumptions break down.
For example, photochemical reactions and the presence of conical intersections (CIs) close to reaction paths necessitate nonadiabatic dynamics beyond the BOA. 
While methodologies to describe these dynamics, such as surface hopping, are readily available, they are approximate and are not broadly applicable~\cite{sholl1998generalized,ben2000ab, subotnik2016understanding,Curchod2018-at,Jain2022-qn,Krotz2022-ea}.
Further, recent work on higher-dimensional features of the PES such as bifurcations and second order saddle points has highlighted their importance in controlling the outcomes of chemical reactions~\cite{Rehbein2011-bk,Hare2017-wm,Pradhan2019-qc,Mirzanejad2023-aw}. 
When these features are present, it is necessary to move beyond the picture of one-dimensional reaction pathways, complicating a theoretical analysis.
While dynamical simulations can account for these challenges, their computational cost and lack of generality are a significant obstacle for a large scale exploration and characterization of the PES.
For example, it remains a major challenge to a priori identify regions of the PES in which the aforementioned features are expected to have a strong influence on reaction dynamics.
In this vein, we propose that recent theoretical developments in the field of topological condensed matter physics can be used to characterize the presence and influence of features such as CIs on reaction outcomes. 
More concretely, we will explore how the concept of topological invariants can connect the structure of the PES with the appearance of nonadiabatic dynamics, and showcase this connection in a model system.  

The paper is structured as follows: We begin by revisiting the BOA approximation and the appearance of nonadiabatic couplings as geometric objects.
Through this perspective, we connect the theory of nonadiabatic couplings with topological invariants encountered in characterizations of condensed matter systems. 
Lastly, we demonstrate our findings using a model system and show how topological invariants constrain the distribution of nonadiabatic couplings along reaction paths.

\section{The Born-Oppenheimer approximation and nonadiabatic couplings}
The PES itself is defined through the eigenvalues of electronic Schrödinger Equation (SE):
\begin{equation}
    \label{eq:SE_elec}
    H_{el} \ket{\phi_i\Rt} = \epsilon_i\Rt \ket{\phi_i\Rt},
\end{equation}
where $\bs{R}(t)$ represents the nuclear coordinates, and $\epsilon_i$ is the instantaneous energy of the $i^{th}$ electronic state $\ket{\phi_i\Rt}$. The nuclear coordinates are explicitly dependent on time $t$, while the electronic wavefunctions and energies are parametrically dependent upon nuclear coordinates. 
Chemical reactions can then be interpreted as the dynamics of the nuclei on this PES. Their dynamics are determined by the solution of the time dependent nuclear SE
\begin{equation}
\label{eq:nuc_SE}
 i\frac{d}{dt} \ket{\Psi\left(\bs{R},t \right) } = H\ket{\Psi\left(\bs{R},t \right) },
\end{equation}
where $H$ is the full Hamiltonian describing electrons and nuclei in atomic units. Going forward, $t$ dependence will be suppressed for brevity. As the electronic eigenfunctions form a complete basis, one can expand the exact solution to Eq.~\ref{eq:nuc_SE} in this basis as
\begin{equation}
    \ket{\Psi\left(\bs{R},t \right) }= \sum_i \chi_i(\bs{R},t) \ket{\phi_i(\bs{R})},
\end{equation}
where $\chi_i(\bs{R},t)$ is the $i$-th nuclear wavefunction. 
Inserting this Ansatz into Eq.~\ref{eq:nuc_SE} and applying some algebra, we obtain the matrix differential equation
\begin{equation} \label{eq:nuclear_dynamicalequation}
   i \frac{d}{dt} \bs{\chi} =  [\frac{1}{2}(\nabla_\bs{R}+\bs{A})^2 +\bs{\epsilon}] \bs{\chi},
\end{equation}
where $\bs{\chi}$ is the vector of nuclear wavefunctions, $\bs{\epsilon}$ is the diagonal matrix of the electronic energies $\epsilon_i$, and $\bs{A}$ is the matrix of nonadiabatic coupling vectors 
\begin{equation}
\label{eq:nac_def}
\AA_{ij}(\bs{R}) = \braket{\phi_i(\bs{R})|\nabla_{\bs{R}}|\phi_j(\bs{R})},
\end{equation}
whose importance will be discussed in details below.
The term
\begin{equation}
 \frac{1}{2}(\nabla_\bs{R} +\bs{A})^2
\end{equation}
is sometimes referred to as the dressed kinetic energy operator, which allows us to interpret the nuclear SE as a non-abelian gauge theory.\cite{Mead1992-og,worth2004beyond,wittig2012geometric}
In this picture, the heavy nuclei move in the potential determined by the fast moving electrons under the influence of the non-abelian gauge field $\bs{A}$. 
The role of the gauge field is to facilitate coupling between different electronic states due to the motion of the nuclei.

The BOA is central to modern quantum chemistry because it admits the assumption that over the course of a reaction a chemical system remains in a single electronic state, the evolution of which is described by
\begin{equation}
 i \frac{d}{dt}\chi_n = [\frac{1}{2}\nabla_\bs{R}^2 +\epsilon_n]\chi_n.
 \end{equation}
The BOA holds for a expansive number of systems due to the large difference between the electron mass and the masses of the nuclei.
It fundamentally relies on a substantial difference in energy between the electronic state under consideration and the remaining electronic states. When this prerequisite is violated and the energy separation between the state of interest and the remaining states dwindles, the BOA may not furnish accurate predictions.

To achieve this simple single state model, certain terms are omitted from the Schrodinger equation, among which are the first derivative nonadiabatic couplings (NACs) defined in Eq.~\ref{eq:nac_def}.
Physically, we can interpret these terms as an interaction between distinct electronic states induced by nuclear motion on the PES\footnote{We omit $i$ in the definition of the  Euler Connection to for it to be real}. 
The Hellman-Feynman Theorem provides a useful relationship between $\AA_{ij}(\bs{R})$ and the energies of the coupled electronic states:
\begin{equation}
\label{eq:NACdefinition_hellfeyn}
        \AA_{ij}(\textbf{R}) = \frac{\braket{\phi_i(\textbf{R})|\nabla_\bs{R} H|\phi_j(\textbf{R})}}{\epsilon_j - \epsilon_i}.
\end{equation}

This representations allows us to consider three distinct regimes in terms of the applicability of the BOA, identifiable by the magnitude of \mbox{$\Delta E = \epsilon_j - \epsilon_i$}.
In the first regime $\Delta E \gg \braket{\phi_1(\textbf{R})|\nabla_\bs{R} H|\phi_2(\textbf{R})}$, in which case the NACs are negligible and the BO approximation may be invoked. In contrast, the next regime features two adjacent electronic states which cross one another at what is commonly known as a conical intersection (CI).\cite{Yarkony2001-xd,domcke2012role,matsika2011nonadiabatic} This corresponds to a singularity in Eq.(\ref{eq:nac_def}) as $\Delta E = 0$, and consequently the BOA breaks down. Reaction dynamics in such a case are radically different from what a kinetic theory relying on a single PES would suggest. Reactions with these features are a focus of photochemistry, and understanding the influence of CIs on the reaction dynamics is a focal point of research in this field.~\cite{bernardi1996potential,Yarkony2001-xd,levine2007isomerization,Curchod2018-at,Matsika2021-um,Boeije2023-xa}.
Finally, there is the regime defined by an intermediate value of $\Delta E$, such that the numerator and denominator in Eq.~\ref{eq:NACdefinition_hellfeyn} are on similar orders of magnitude. In this case, NACs cannot necessarily be ignored and the dynamics depend on the energetic details of the system and the distribution of the NACs~\cite{Malhado2014-aj}.
In practice, exact calculations of the dynamics are not feasible due to the computational cost. As a result, there are several popular approximations that can be interfaced with \textit{ab-initio} quantum chemistry codes to describe the dynamics, such as surface hopping and \textit{ab-initio} multiple spawning~\cite{Mai2018-ve,Curchod2018-at}. 
However, analyzing dynamics remains challenging due to their complexity and the computational cost, in particular for larger molecules~\cite{Li2022-fh}.

\section{Nonadiabatic couplings as geometric objects}
In this section we wish to connect NACs with geometric and topological concepts derived from the electronic wavefunctions, as commonly encountered in condensed matter physics.
The NACs are generally organised in a matrix, in which the diagonal components are purely imaginary, while off-diagonal components are not constrained.
It is well known, that the diagonal components are connected to a geometric phase factor $e^{i\gamma_n}$, where $\gamma_n$ is known as the geometric phase in chemistry, and as the Berry phase in condensed matter physics. The phase is expressed as:
\begin{equation}
    \label{eq:geom_phase}
   \gamma_n = \frac{1}{2\pi} \int_C d\RR \cdot \AA_{nn}(\RR),
\end{equation}
where the integral is taken along a path $C$ on the PES.
This phase is a geometric quantity, because its value only depends on the choice of path. Additionally, the integral itself is not gauge invariant, which complicates its interpretation and numerical computation.
Its relevance stems from the fact that this phase can give rise to interference effects between paths with different values of $\gamma_n$, which in some cases radically alters the dynamics.~\cite{ryabinkin2017geometric}

When the path $C=\partial S$ is chosen to be a closed loop that encloses an area $S$, the geometric phase 
allows for the definition of a topological invariant, i.e. its value is quantized to integers and is independent of the choice of path and gauge. 
The topological invariant is the Chern Number $c_n$, which counts the number of level crossings enclosed by the loop $\partial S$.
\begin{equation}
    \label{eq:chern}
   c_n = \frac{1}{2\pi} \int  \nabla\times \AA_{nn} (\RR) \cdot d\bs{S} - \frac{1}{2\pi} \int_{\partial S}  \AA_{nn}(\RR) \cdot d\bs{C}  \in \mathrm{Z}
\end{equation}
where the integral is taken along a path $\partial S$ on the PES, $S$ is the surface bounded by the path. The above definition can be applied to both an open region with a boundary and a closed region with no boundary.

In condensed matter physics, the diagonal component $\AA_{nn}$ is called the U(1) Berry Connection and its curl $\Omega_{n}=\nabla\times \AA_{nn}$ is the gauge independent Berry Curvature~\cite{xiao2010berry}.
The Berry phase and Berry Curvature of electronic wavefunctions play an essential role in understanding a variety of physical phenomena, including charge polarization, orbital magnetism, various Hall effects, and topological charge pumping in condensed matter physics~\cite{Resta2000-kz,xiao2010berry,Andrei_Bernevig2013-nf}. Particularly, the Chern Number can be directly related to the quantized Hall conductance observed in the quantum Hall effect.\cite{xiao2010berry} 

In recent years, generalizations of the Chern number have revolutionized our understanding of electronic states in condensed matter physics~\cite{Moore2010-sd}. 
In the framework of topological band theory, the crossings between energy levels of electrons are characterized by topological invariants, which predict the presence of so-called topological boundary states~\cite{Moore2010-sd,Bansil2016-wb}. Recent efforts aim to topologically characterize all possible electronic states in crystalline materials based on high throughput calculations, estimating that approximately 53\% of known crystalline materials may possess topologically protected electronic states~\cite{Wieder2021-fc,Vergniory2022-cj}. 

This begs the question if these new concepts could similarly be applied to problems in reaction chemistry, as crossings of energy levels happen frequently on the PES, e.g. at CIs~\cite{Mead1979-ox}.
However, the distinct quantum mechanical descriptions of solids versus molecules makes it difficult to directly apply them to molecular problems. For example, molecules are zero-dimensional objects without boundary and therefore cannot possess boundary states. Further, the Hamiltonians describing crystalline materials are formulated in reciprocal space, complicating a direct generalization of these   methods to a molecular setting which is formulated in real space.
In the following, we discuss how these issues can be overcome. We introduce a framework that allows us to connect topological invariants with the distribution of NACs, suggesting that topological methods classify the nonadiabaticity of chemical reactions.

For the Chern Number and geometric phase factor to be non-zero, time-reversal symmetry (TRS) needs to be broken.
This stems from the fact TRS allows the choice of a real gauge, which enforces the diagonal NACs to vanish as they are purely imaginary.
On a Hamiltonian level, this is represented by a reality condition $H_{el}^*(\RR)=H_{el}(\RR)$. As many chemical reactions do not break TRS, the Chern Number has attracted little attention in the context of chemical reaction dynamics.

In contrast, the off-diagonal components $\AA_{ij}$ with $i \neq j$ are not constrained by TRS, and generally take on non-zero values. 
It is therefore desirable to explore generalizations of the Chern Number, that are defined in terms of the off-diagonal NACs, which are of direct relevance to the discussion of nonadiabatic dynamics in chemical reactions.
The Euler Class is such a generalization~\cite{Ahn2018-ux,ahn2019failure,bouhon2020non,Song2021-af}.
It has recently been the subject of intense studies in the context of fragile topology in condensed matter physics for materials such as twisted bilayer graphene and other low-dimensional compounds~\cite{Po2019-am,Song2020-cd}. In analogy to the Chern Number, the Euler Class $\chi$ is an integer number, and is defined via the off-diagonal NACs between two electronic states $\ket{\phi_i}$ and $\ket{\phi_j}$ as 
\begin{equation}
\label{eq:euler_def}
\chi_{ij} =  \frac{1}{2\pi} \int_S \nabla_{\mathbf{R}} \times  \bs{\mathrm{A}_{ij}}(\bs{R}) \cdot d\mathbf{S} -  \frac{1}{2\pi}  \oint_{\partial S} \mathrm{\AA}_{ij}(\bs{R})\cdot d\bs{C} .
\end{equation}
One often defines the Euler Form $\mathrm{Eu}_{ij}(\bs{R}) = \nabla_{\mathbf{R}} \times  \bs{\mathrm{A}}_{ij}(\bs{R})$ in analogy to the Berry Curvature. The integral of $\mathrm{Eu}_{ij}(\bs{R})$ is performed in a two-dimensional (2D) region $S$ in the parameter space $\RR$, while the loop integral of $\AA_{ij}$ is performed at the boundary $\partial S$ of the region $S$. The above definition can be applied to both an open region with a boundary and a closed region with no boundary.~\footnote{In contrast to the Chern Number, which can be defined for any number of states, the definition of the Euler Class for systems with more than three states is more involved. If the three states are separated by a gap from the other states, our definition here is sufficient, and we restrict our discussion to this case.}

The Euler Connection, Euler Form and Euler Class are in analogy to the Berry Connection, Berry Curvature and Chern Number, but the former are only well-defined for a pair of real states. This reality condition turns out to be much easier to satisfy in chemical reactions, as compared to the electronic states in crystals. In crystals, the crystal Hamiltonian $H(\kk)$ of electrons is a function of the momentum $\kk$, which reverses its sign under time reversal $T$. Thus,  time reversal requires $H(\kk)=H^*(-\kk)$, which does not guarantee $H(\kk)$ to be real. 
To enforce a real Hamiltonian, either $PT$ symmetry, where $P$ is spatial inversion, or $C_{2}T$ symmetry, where $C_2$ is a two-fold rotation, is required for the Hamiltonian to satisfy $H^*(\kk)=H(\kk)$, which allows the choice of a real gauge needed to define the Euler Class~\cite{Ahn2018-ux,ahn2019failure}. 
In contrast, the Hamiltonian $H(\RR)$ describing a chemical reaction is defined over the space of nuclear coordinates $\RR$, which are even under TRS. 
In molecular systems, TRS is broken by external magnetic magnetic fields, circularly polarized electric fields or in molecules with non-zero spin, e.g. radicals or transition metals. 
As many chemical reactions do not include these effects, the Euler Class is expected to be generally well defined. 

The purpose of this paper is to establish the utility and applicability of modern topological concepts such as the Euler Class in the field of chemical reaction dynamics.
To that end, we propose that the significance of the Euler Class lies in its ability to diagnose and categorize the distribution of NACs in the region $\bs{S}$, as different values of the Euler Class require manifestly different distributions of NACs.
Therefore, Eq.~\ref{eq:euler_def} is the central equation to our discussion, as it explicitly connects nonadiabatic couplings with topology.

In contrast to the Chern Number, which measures the number of individual crossings in surface $\bs{S}$, the Euler Class measures the stability of individual \textit{pairs} of CIs contained in $\bs{S}$. An Euler Class of $\chi = 0$ indicates there are either no CIs contained in $\bs{S}$ or that the CIs in $\bs{S}$ can be removed by changing parameters in the Hamiltonian such that they coalesce and the associated states become gapped. 

It is well established, that a CI can be viewed as a source of nonadiabatic couplings. 
Traditionally, their contribution to the reaction dynamics is considered individually. As we will show below, such a local picture does not provide global information about the distribution of NACs.
Rather, one must consider whether the contributions from multiple CIs are constructive or destructive. The question then becomes: are all CIs the same?
The Euler Class allows us to answer this question rigorously. In the following, we will explore these points in detail for a model system.

\subsection{Defining the Model System}
\begin{figure}[ht]
 \includegraphics[width=1.0\columnwidth]{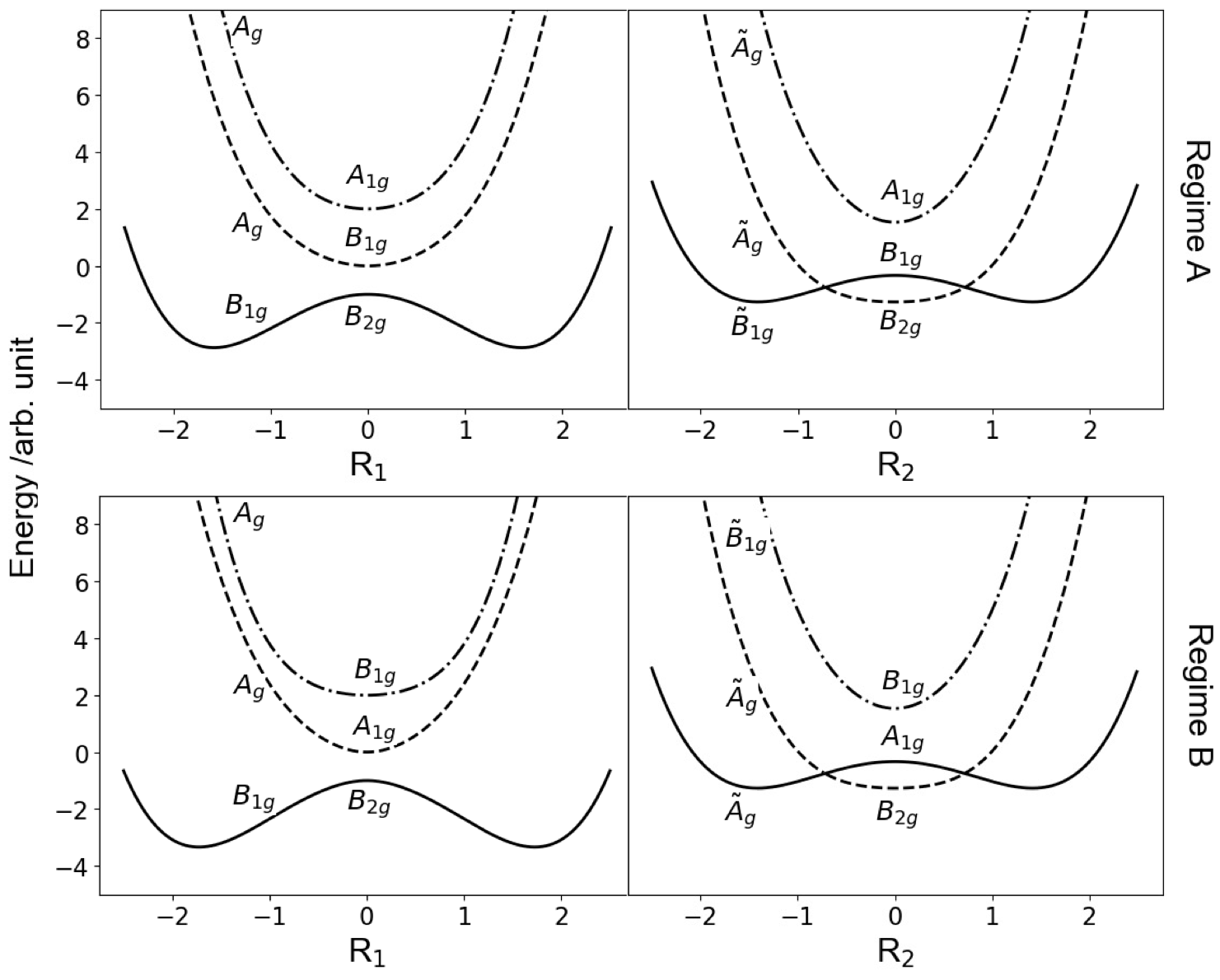}
\caption{
PES plots for two distinct parameter regimes of the model Hamiltonian given in the main text. Energies are displayed for each regime along the high-symmetry axes $R_1$ and $R_2$. The irreducible representations of states corresponding to each energy level are displayed at $\bs{R}_0$ as well as away from $\bs{R}_0$ for each axis respectively
}
\label{fig:PESFig}
\end{figure}
We begin by introducing a 3-state model Hamiltonian, which is a function of two nuclear coordinates $\bs{R} = (R_1,R_2)$, which we interpret as nuclear vibrational modes of the molecule at $\bs{R} = 0$. 
The vibrational modes and irreducible representations (irreps) of the eigenstates of the Hamiltonian stand in analogy those arising in the isomerization of cyclobutadiene (CBD), however couplings between the states differs from those of CBD.
We stress the point that this model does not attempt to accurately model the isomerization of CBD. Instead, we use the key features of this reaction to build a suitable basis for our often abstract discussion.

At $\bs{R_0} = \bs{0}$, the Hamiltonian possesses $D_{4h}$ symmetry and is inspired by the high energy antiaromatic configuration of CBD~\cite{Bosse2021-es}.
Orienting our perspective with respect to this point, the PES along $R_1$ for $R_2 = 0$ resembles the PES of the isomerization in that there are two local minima connected through a first order saddle point. 
Continuing this analogy, we introduce a motion $R_1$ that transforms as the $B_{1g}$ representation of $D_{4h}$.  It corresponds to an energy lowering stretching of the $D_{4h}$ symmetric geometry at $\bs{R}= 0$, which lowers the symmetry of molecule to $D_{2h}$.
To include the presence of CIs and discuss their influence, we introduce a 
second low energy motion along a second coordinate $R_2$ that transforms as the $B_{2g}$ representation.
In analogy to the rhomboidal stretching mode of CBD, this mode increases the energy of the molecule close to $\bs{R_0}$ and lowers the symmetry from $D_{4h}$ to $D_{2h}$.
The basis states of our model are chosen to have same irreps as the three lowest lying singlet states of CBD at the $D_{4h}$ geometry, namely the $A_{1g}$, $B_{1g}$ and $B_{2g}$ irreps. 
These three irreps are generally expected in the case of degenerate molecular orbitals belonging to the $E$ representation. Assuming two electrons in this set of orbitals, the possible symmetry adapted many-electron states are $E \otimes E = A_{1g} \oplus B_{1g} \oplus B_{2g} \oplus A_{2g}$, of which the $A_{2g}$ state is a triplet $S = 1$.
In the basis of singlet states, the Hamiltonian is given as
\begin{equation}
\mathcal{H}(\bs{R}) = \\
\begin{bsmallmatrix} 
a_{11} + b_{11}R_1^2 + c_{11}R_2^2 + g_{11}(R_1^4+R_2^4) & d_{12}R_1 & e_{13}R_1R_2\\
d_{12}R_1 & a_{22} + b_{22}R_1^2 + c_{22}R_2^2 + g_{22}(R_1^4+R_2^4) & f_{23}R_2\\
e_{13}R_1R_2 &f_{23}R_2 & a_{33} + b_{33}R_1^2 + c_{33}R_2^2 + g_{33}(R_1^4+R_2^4)
\end{bsmallmatrix},
\end{equation}
in which the values of parameters (e.g. $a_{11}$, etc) are listed in the Tab. \ref{table:parameters}. 

\begin{table}
\begin{center}
\begin{tabular}{ |c|c|c|c|c|c| } 
\hline
Parameter & Regime A & Regime B & Parameter & Regime A & Regime B \\
\hline
$a_{11}$ & 0 & 0 & $g_{11}$ & .25 & .25 \\ 
$a_{22}$ & 2 & -1  & $g_{22}$ & .8 & .30\\ 
$a_{33}$ & -1 & 2  & $g_{33}$ & .30 & .8\\
$b_{11}$ & 2 & -1.5  & $d_{12}$ & .5 & 2\\
$b_{22}$ & 1 & 2  & $d_{21}$ & .5 & 2\\
$b_{33}$ & -1.5 & 1  &$ e_{13}$ & 2 & 2\\
$c_{11}$ & -1 & -1  & $e_{31}$ & 2 & 2\\
$c_{22}$ & 2 & 2  &$f_{23}$ & 2 & 2\\
$c_{33}$ & 2 & 2  & $a_{32}$ & 2 & 2\\
\hline
\end{tabular}
\caption{Table displaying values of Hamiltonian parameters for the two topologically distinct regimes}
\label{table:parameters}
\end{center}
\end{table}

There are three symmetries ($C_4, C_2,C_2'$) whose action on the Hamiltonian will be important for the following discussion. The action of the symmetries on the Hamiltonian are expressed through the following commutation relations:
\begin{equation}
\begin{split}
\left[ \mathcal{H}(\bs{R}), D(C_4)\right] &= \mathcal{H}(-\bs{R})\\
\left[ \mathcal{H}(\bs{R}), D(C_2)\right] &= \mathcal{H}(R_1,-R_2) \\ 
\left[ \mathcal{H}(\bs{R}), D(C_2')\right] &= \mathcal{H}(-R_1,R_2)
\end{split},    
\end{equation}
where $D(C_4)$ ($D(C_2)$, $D(C'_2)$) are the matrix representations for $C_4$ ($C_2$, $C'_2$) symmetry respectively. A detailed derivation and further details are provided in App.~\ref{app:reps}.
These commutation relations constrain the form of the Hamiltonian and break down the symmetry group of the Hamiltonian from $D_{4h}$ to $D_{2h}$ along to two high-symmetry lines $R_1 =0$ and $R_2 = 0$. Along these lines we can label the eigenstates of the Hamiltonian by the irreps of $D_{2h}$. Because $R_1$ and $R_2$ correspond to distinct $D_{2h}$ symmetric geometries, irreducible representations with the same label are not necessarily the same along these two lines. Therefore we adopt the convention of labeling irreps along $R_1 = 0$ with a tilde, e.g. $A_{1g}$ vs $\tilde{A}_{1g}$. 

In the following discussion, we will study two regimes of parameters. 
For both regimes the model possesses almost indistinguishable PESs along the high-symmetry lines, but are distinguished through a topological invariant, the Euler Class.
Figure~\ref{fig:PESFig} compares the PESs along high-symmetry lines for each regime according to the irreps and energies of each state. The key features in both parameter regimes are two minima along $R_1$ representing stable chemical structures, which are connected by an energetic maximum along $R_1$ (a saddle point in $\RR$ space) corresponding to a transition state at $\bs{R}_0$. In the context of our putative reaction, this path corresponds to the minimum energy pathway. Along $R_2$, there are two CIs (nodal crossings), with $\bs{R}_0$ lying at the midpoint between them. We will show that despite lying away from the proposed reaction pathway, these structures play a central role in determining nonadiabatic couplings along the reaction path.

\subsection{Topological Classification of the two regimes}
\begin{figure}[ht]
 \includegraphics[width=0.8\columnwidth]{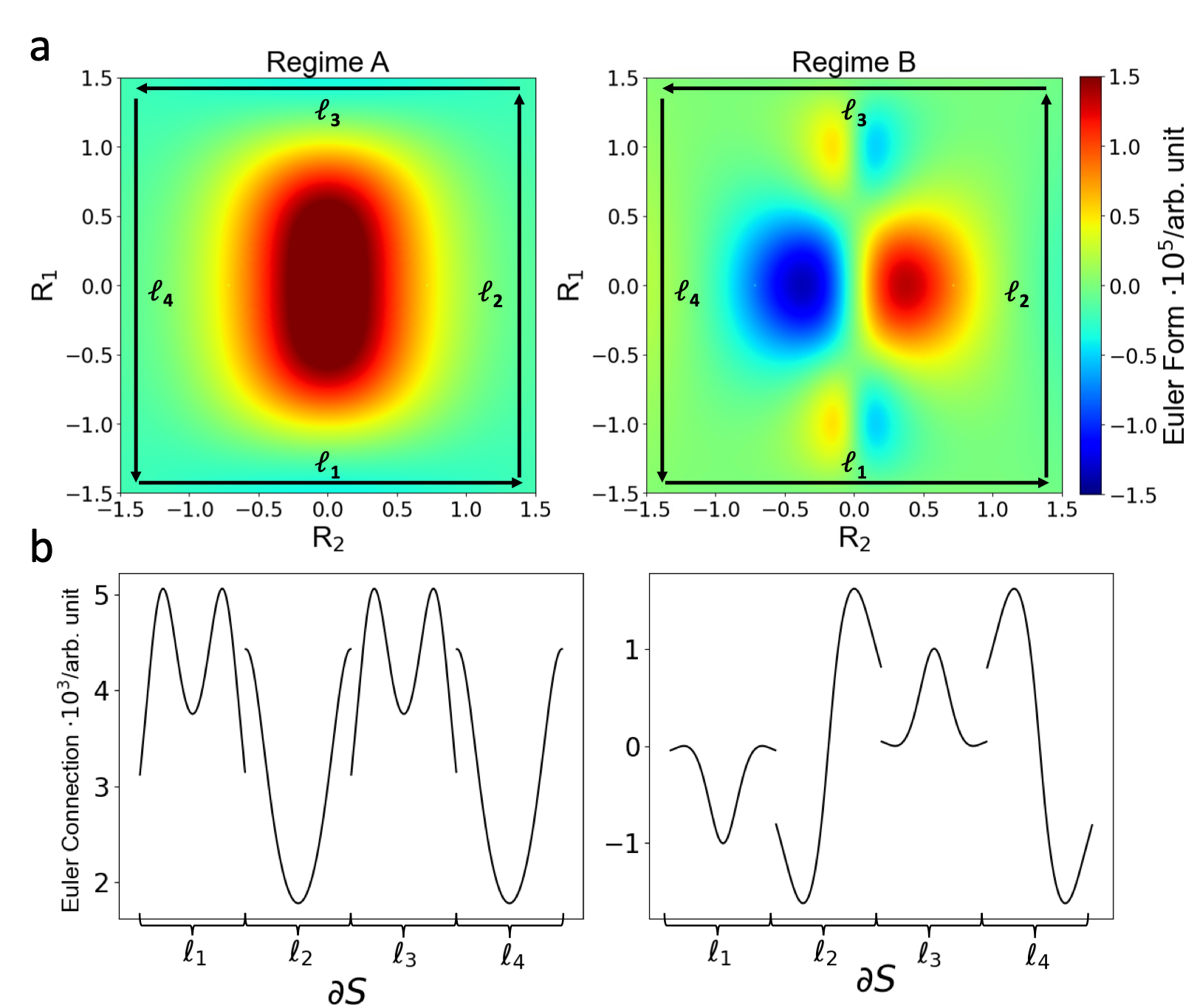}
 \caption{(a) Color plots of the Euler Form over a square region for both parameter regimes. The Euler Form for regime A is symmetric with respect to both coordinate axes, indicating a nonvanishing contribution to the Euler Class. The Euler Form for regime B is symmetric with respect to the $R_1$ axis and anti-symmetric with respect to the $R_2$ axis resulting in no contribution to the Euler Class. In both plots, the arrows labeled $\ell_i$ represent segments of the boundary used in the calculation of the Euler Connection. 
 (b) Plot of the Euler Connection over the boundary of the region shown in (a). The symmetry properties of the Euler Connection for each regime match those of the Euler Form. The discontinuities are due to the choice of path, which project out only one component along each segment $\ell_i$.}
 \label{fig:ECSymm}
 \end{figure}

An essential feature of our model is the fact that the PES is virtually identical for our two distinct parameter regimes.
In the context of the BOA, in which dynamics are solely determined by energy profiles of the PES, one would expect that the dynamics along the putative reaction pathway connecting the two minima are the same.
In addition, there would be no possibility to distinguish the two regimes from each other in the framework of TST, as the mininum energy pathways would be identical.

Below, we will show that the two regimes are fundamentally different from each other. They are distinguished by different values of the Euler Class $\chi$ defined over the two lowest energy states, which corresponds to a categorically different distribution of NACs. As a central result, we will also show that nonadiabatic effects cannot be dismissed for pathways proceeding along the putative reaction coordinate in one regime, while they can be neglected in the other.

For parameter regime A as defined in Table~\ref{table:parameters}, we find an Euler Class $\chi = 1$. 
To rationalize this value, Fig.~\ref{fig:ECSymm} plots the distribution of the Euler Form and Euler Connection for regime A as a function of $\bs{R}$ computed by numerical diagonalization of the Hamiltonian on a square region $S$ centered around $\bs{R} = 0$ (See App.~\ref{app:euler_comp} for details).
The Euler Form displays a clear symmetry $\mathrm{Eu_{12}}(R_1,R_2) = \mathrm{Eu_{12}}(-R_1,R_2)$ and $\mathrm{Eu_{12}}(R_1,R_2) = \mathrm{Eu_{12}}(R_1,-R_2)$ with respect to both the coordinate axes~\footnote{These symmetry relations can be derived using the symmetries of the Hamiltonian. We reserve a detailed discussion for a future publication, as it would be beyond the scope of the paper.}. The the Euler Connection $ \mathrm{\AA}_{12}(\bs{R})$ displays the same symmetry properties as the form along the boundary of domain of integration. Due to these functional forms, the integrals in Eq.~\ref{eq:euler_def} may be non-zero, and in the case of our model, result in $\chi = 1$.
\begin{figure}[ht]
 \includegraphics[width=0.9\columnwidth]{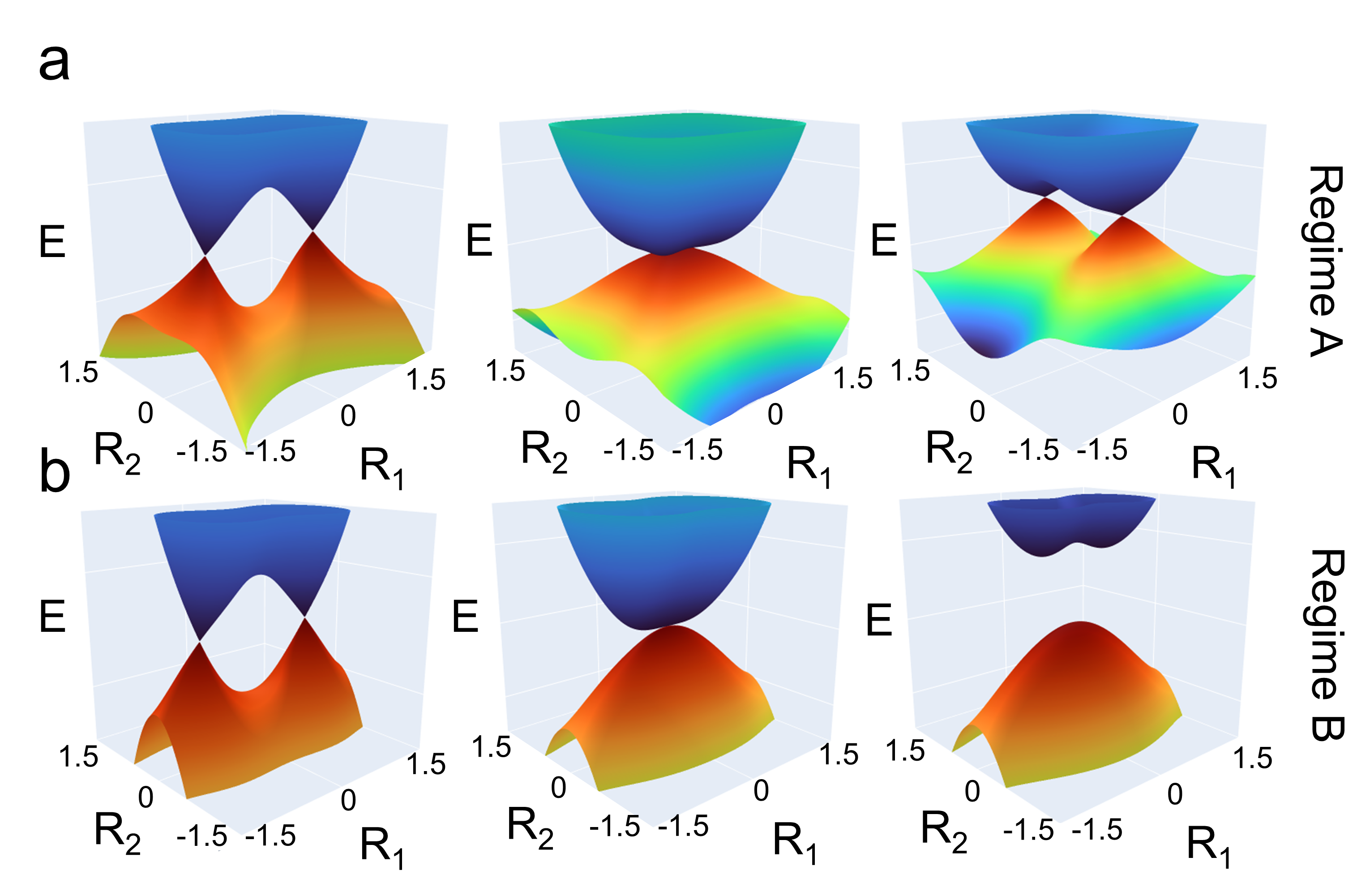}
\caption{Fate of CIs in systems with differing Euler Classes. (a) In a system with $\chi = 1$ a smooth change in parameters ($a_{33}$ varies from $-1$ to $1$) first brings the CIs along the $R_2$ axis, after which the electronic states fail to separate, and the CIs are maintained along the $R_1$ axis at new coordinates. (b) In a system with $\chi = 0$ under a smooth change in parameters ($a_{11}$ varies from $-1$ to $1$), the CIs are first brought together along the $R_2$ axis, and then eliminated, with the electronic states becoming well-separated. }
\label{fig:AnniFig}
\end{figure}
As previously discussed, the significance of the Euler Class from a topological perspective is to describe the robustness of the pair of CIs with respect to smooth parameter changes of the Hamiltonian that preserve TRS.~\footnote{The general definition of smooth parameter changes can become complicated. For the purpose of this paper, we consider a parameter change smooth if it does not introduce additional crossings between energy levels, while preserving the symmetries of the Hamiltonian.}
A non-zero Euler Class implies that the CIs will fail to annihilate if they are brought together through these parameter changes.

To exemplify this behaviour, we follow the evolution of the CIs as we continuously change the value of $a_{33}$ from $-1$ to $1$.
During the course of this evolution, we find a convergence of the two CIs into a single degenerate point along $R_2$, followed by separation of the pair of CIs along $R_1$ [Fig.~\ref{fig:AnniFig}(a)].
Because of this, we refer to the degeneracies as topologically protected from annihilation. 

Parameter regime B as defined in Table~\ref{table:parameters} shows a sharp contrast with the previous scenario. In this system we find an Euler Class $\chi = 0$.
This value can be rationalized by the distribution of the Euler Class and Form as a function of $\bs{R}$. Specifically, the Euler Form is anti-symmetric with respect to the $R_2$ axis $\mathrm{Eu_{12}}(R_1,R_2) = -\mathrm{Eu_{12}}(-R_1,R_2)$, and symmetric with respect to the $R_1$ axis $\mathrm{Eu_{12}}(R_1,R_2) = \mathrm{Eu_{12}}(-R_1,R_2)$ resulting in an overall antisymmetric distribution. While the values of the Euler Form are generally non-zero, the overall antisymmetry results in a cancellation during the integration in Eq.~\ref{eq:euler_def}, and both integrals vanish identically.
As $\chi = 0$, it should be possible to annihilate the two CIs by a suitable smooth change of parameters.
Changing the value of $a_{22}$ from -1 to 1, we observe convergence of the two CIs along $R_2$ into a single degenerate point followed by a gapping of energy levels [Fig.~\ref{fig:AnniFig}(b)].
Therefore, we refer to regime B as topologically trivial, which implies it is equivalent to a model without CIs from a topological perspective.

\subsection{Distribution of NACs and Connection to Euler Class}
\begin{figure*}[ht]
 \includegraphics[width=0.9\columnwidth]{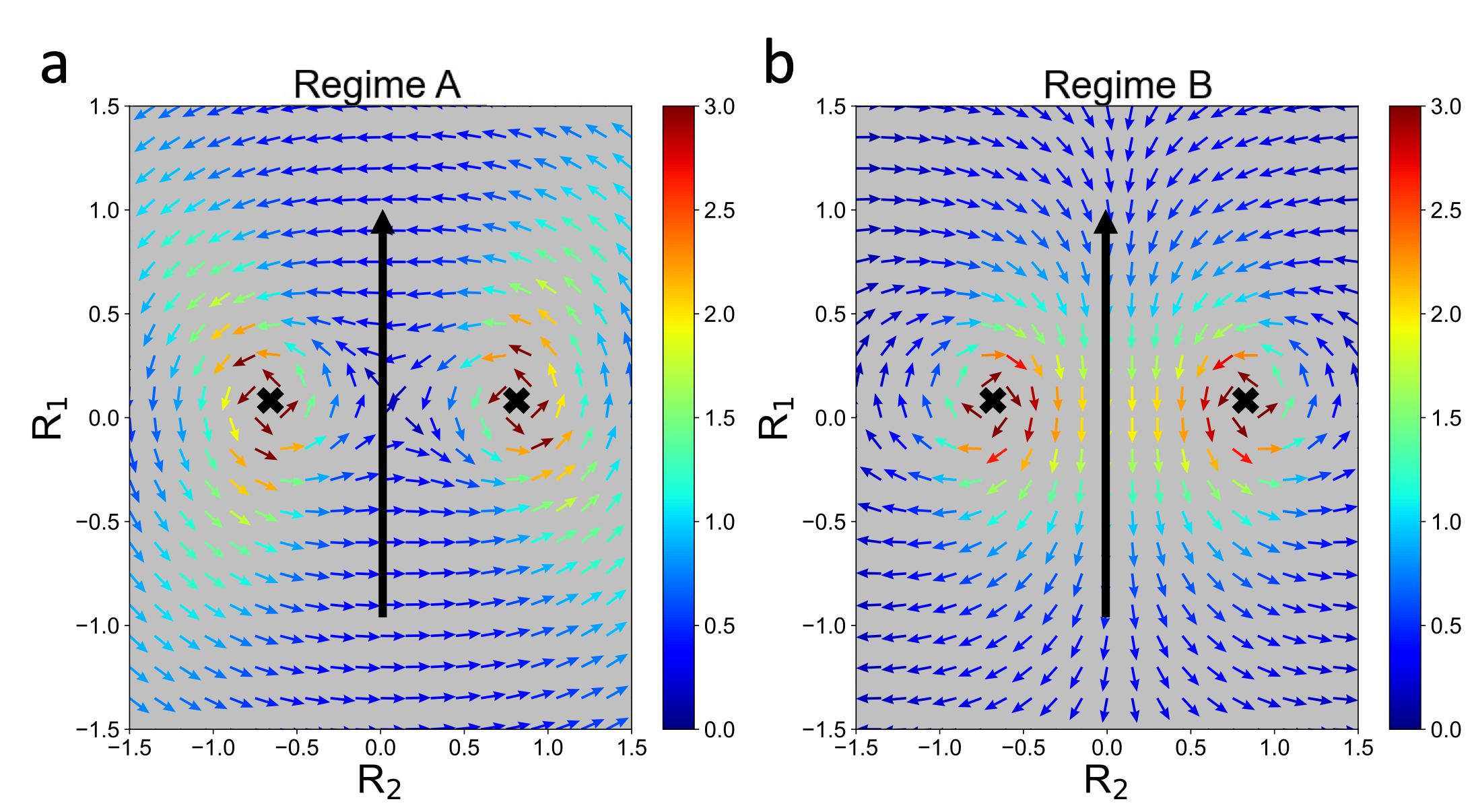}
\caption{Vector field plots of $\AA_{12}(\bs{R})$ for the topologically distinct regimes, in which the colors indicate the norm of the NACs. Black arrows denote the putative reaction path, and each CI is denoted with a black {\bf x}. (a) An Euler Class $\chi = 1 $ corresponds to a distribution of NACs that is characterized by a vanishing norm of NAC vectors close to the minimum energy pathway. (b) An Euler Class $\chi = 0$ is characterized by a distribution of NACs in which the norm of the NAC vectors are non-negligible along the reaction path.}
\label{fig:NAFig}
\end{figure*}
What is the connection between the ability to gap out a pair of CIs and the distribution of NACs in their vicinity?
To understand their connection, we plot the NACs for both parameter regimes as a function of $\bs{R}$.
Starting with regime A, Fig.~\ref{fig:NAFig} (a) shows that the coupling vectors are large in magnitude close to the CIs and away from the proposed reaction path, however their magnitude is small in the region containing it. 
The distribution of nonadiabatic couplings around each CI is vortex-like, i.e. the vectors follow a counterclockwise rotation around both CIs. 
The total NAC distribution can be approximated from summing up the individual contributions of each CI.
At $\bs{R_0}$, the summation results in a cancellation of NACs.
In proximity to $\bs{R_0}$ and to the reaction path the NACs are small in magnitude, and the components along the $R_1$ direction vanish. 
In contrast, the distribution of NACs for regime B is markedly different [Fig~\ref{fig:NAFig}(b)], i.e. the NAC-vector rotates counterclockwise around one CI but clockwise around the other CI. 
Due to the opposite orientation of the vortices, there is an addition of NACs along the $R_1$ direction, while the components along the $R_2$ direction vanish.
This results in NACs that are large in magnitude both near the CI, as well as across the region that the proposed reaction path must traverse.

Canonically, the magnitude of the NACs is rationalized by the energy difference between the coupled states $\ket{\phi_1}$ and $\ket{\phi_2}$, as already discussed for Eq.~\ref{eq:NACdefinition_hellfeyn}.
In the case of our model, this analysis is not conclusive, since both PES are almost identical energetically, yet the distribution of NACs differs markedly.
\begin{figure}[t]
\includegraphics[width=0.55\columnwidth]{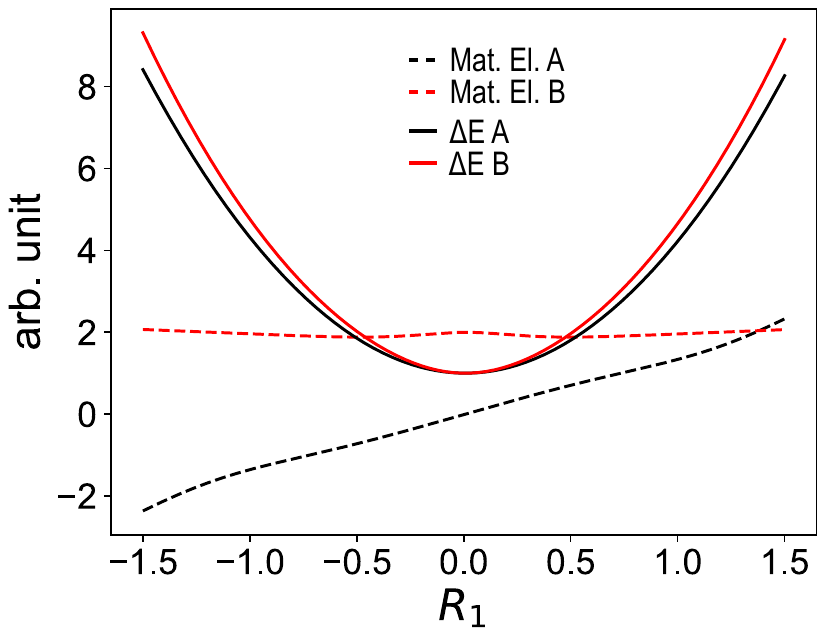}
\caption{Plot of non-zero matrix elements and energies for two regimes along $R_1$. The matrix elements are defined as $\braket{\phi_i(\textbf{R})|\nabla_\bs{R} H|\phi_j(\textbf{R})}$. For regime A, only the $R_2$ component is non-zero, whereas only the $R_1$ component is non-zero in regime B. Though the difference in $\Delta E$ between two regimes is extremely small, there are qualitative differences in the NACs [Fig.~\ref{fig:NAFig}]. These are due to a sign-change of the matrix element for regime A, while there is none for regime B. 
}
\label{fig:mat_el}
\end{figure}
Fig.~\ref{fig:mat_el} compares the matrix elements and the energy difference between the two models along the reaction path $\bs{R} = (R_1,0)$.
As expected, the denominators in Eq.~\ref{eq:NACdefinition_hellfeyn} are virtually identical, however we find stark differences in the matrix elements. This highlights the importance of exactly calculating the nonadiabatic couplings, instead of relying on approximate methods that neglect matrix element effects, e.g. as in the commonly used Zhu-Nakamura variant of surface hopping~\cite{Zhu2001-by}.

These findings suggest that protection of the CIs from annihilation in the topological sense implies protection of adiabaticity in the chemical sense, at least for our model system. 
For regime A, we hypothesize that the reaction dynamics may solely determined by the barrier height along the minimum energy pathway. As the strong reduction in magnitude of the NACs along this pathway is not fine-tuned and does not explicitly depend on energy separations or other parameters, we expect this phenomenology to be general as long as $\chi = 1$. 
In contrast, the contribution of NACs to the dynamics in regime B will depend on the details of the Hamiltonian, e.g. the energy gap between electronic states and the separation of the CIs along $R_2$.

\section{Conclusion}
In summary, we have developed a model for a simple chemical reaction that is inspired by the isomerization of cyclobutadiene. The PES of this model possesses two minima which are connected along a minimum energy pathway, which is situated between two CIs.
A key feature of this model is the existence of two parameter regimes that feature almost identical PESs, yet a distinct distribution of NACs. 
In one case, regime A, the reaction is predicted to proceed adiabatically, since the NACs along this path vanish.
In the other regime, the NACs do not vanish, and, depending on their magnitude, may influence the dynamics strongly.
We have shown that one can capture this behaviour through the Euler number, a topological invariant that depends on the distribution of NACs.
The topological analysis suggests that the global NAC distribution can be approximated from summing up the individual contributions of each CI in terms of magnitude and orientation.
In regime A, the orientations of NACs around the CIs are the same, resulting in a drastic reduction of the mangnitude of NACs along the minimum energy pathway.
In contrast, the orientations of NACs around the CIs are opposite in regime B, which results in an accumulation of NACs along the minimum energy pathway.
 
The goal of this perspective has been to establish a connection between the topological language of invariants and nonadiabatic reaction dynamics. The connection between the distribution of NACs and the values of the Euler Class suggested here are a first step in this direction, as we have explicitly shown and analyzed how such a connection arises in our model system. It still remains to be explored how general this connection is. 

It also remains to be shown that the dynamics indeed proceed as predicted here and if approximate methods such as surface hopping can capture the distinction between these regimes.
For example, the Zhu-Nakamura variant of surface hopping approximates NACs based only on energetic features of the PES. As we have shown above, the main distinction between the two regimes lies not in the energetics, but in the structure of the matrix elements appearing in the expression for the NACs. 

We have omitted a detailed discussion about the relation between spatial symmetries and the distribution of NACs. 
This is due to the fact that the topological quantities discussed here only rely on the presence of time reversal symmetry, i.e. the results discussed here would remain true even if the spatial symmetries were to be weakly broken.
Spatial symmetries, however, can greatly simplify the analysis, as it is possible to connect the topological properties with the symmetry eigenvalues of the electronic states involved. 
The connection between symmetries and topology has been a focal point of research in condensed matter physics and has enabled a systematic search of materials with topological features in their electronic structure~\cite{Wieder2021-fc,Vergniory2022-cj}.
It begs the question if such a theory could similarly enable a systematic screening for reactions in which the phenomenology discussed here would apply, i.e. cases with non-zero Euler number.

We note that the analysis discussed here is also applicable to pairs of excited states, and is therefore relevant to problems in photochemistry.
In photochemical problems, the dynamics take place on excited state PESs and CIs serve as radiationless relaxation pathways lower energy surfaces.
However, in realistic calculations one commonly observers transitions to lower energy surfaces in a wide region around the CIs, rather than at the point of energy degeneracy. 
These transitions are facilitated by large NACs surrounding the CIs, which often is rationalised by small energy gap between states near the CI. 
Yet, as we have shown, additional complexity arises in cases with complex PES, for which only considering energetic properties can be misleading.
Gaining a comprehensive understanding of the distribution of NACs beyond simple energetic pictures could therefore not only help to rationalize the results of dynamical calculations and possibly experiments, but could also provide new guidelines on how to manipulate the dynamics by changing the topological properties of the system. 


\bibliography{lit}

\newpage

\appendix
\section{Hamiltonian Construction From Point Group Symmetries}
\label{app:reps}
The basis functions of the 3-level model Hamiltonian presented in this work transform as the $A_{1g}$, $B_{1g}$ and $B_{2g}$ irreducible representations of the $D_{4h}$ point group.
In this basis, the matrix representations for the generators of the $D_{4h}$ point group, namely four-fold rotation $C_4$, two-fold rotation $C'_2$ and inversion $I$, are given as:
\begin{equation}
\begin{split}
D(C_4) = 
\begin{bmatrix}
-1 & 0 & 0 \\ 
0 & 1  & 0 \\ 
0 & 0 &-1 
\end{bmatrix} \ \ 
D(C_2') = 
\begin{bmatrix}
-1 & 0 & 0 \\ 
0 & 1  & 0 \\ 
0 & 0 & 1 
\end{bmatrix} \ \
D(I)  = 
\begin{bmatrix}
1 & 0 & 0 \\ 
0 & 1  & 0 \\ 
0 & 0 & 1 
\end{bmatrix}
\end{split}.
\end{equation}
For nonzero $\bs{R}$, the  $D_{4h}$ symmetry of the Hamiltonian is broken. By construction the the Hamiltonian will possess $D_{2h}$ symmetry along the $R_1$ and $R_2$ axes.  There are two distinct ways to break down $D_{4h}$ to $D_{2h}$, which we implement along the $R_1 = 0 $ and $R_2 = 0$ lines respectively. For all other values of $\bs{R}$ the Hamiltonian does not belong to any point group.
Along the $R_1$ axis the group is $D_{2h}$ in the canonical orientation. By subduction of symmetry the basis functions transform as $A_{1g}\rightarrow A_{g}$, $B_{1g} \rightarrow A_g$, and $B_{2g} \rightarrow B_{1g}$. The group generators are
\begin{equation}
\begin{split}
D(C_2)  = D(I) =  
\begin{bmatrix}
1 & 0 & 0 \\ 
0 & 1  & 0 \\ 
0 & 0 & 1 
\end{bmatrix} \ \
D(C_2')  = 
\begin{bmatrix}
1 & 0 & 0 \\ 
0 & 1  & 0 \\ 
0 & 0 & -1 
\end{bmatrix}
\end{split}
\end{equation}
Along the $R_2$ axis the point group is $\tilde{D}_{2h}$. By subduction of symmetry, the basis functions transform as $A_{1g}\rightarrow \tilde{A}_{g}$, $B_{1g} \rightarrow \tilde{B}_{1g} $, and $B_{2g} \rightarrow \tilde{A}_{g}$\footnote{To distinguish the irreps along $R_1$ and $R_2$, we denote the latter with a tilde.}. The group generators are
\begin{equation}
\begin{split}
\tilde{D}(C_2) = \tilde{D}(I) =  
\begin{bmatrix}
1 & 0 & 0 \\ 
0 & 1  & 0 \\ 
0 & 0 & 1 
\end{bmatrix} \ \
\tilde{D}(C_2') = 
\begin{bmatrix}
1 & 0 & 0 \\ 
0 & -1  & 0 \\ 
0 & 0 & 1 
\end{bmatrix}
\end{split}
\end{equation}
The constraints placed on the Hamiltonian by the given symmetry requirements are expressed through a set of commutation relations for each symmetry element $\bs{g}$ of $D_{4h}$
\begin{equation}
  D(g) \mathcal{H}(\bs{R}) D(g)^{\dagger} =   \mathcal{H}(\bs{gR}).
\end{equation}
For example the action of $C_4$ is given as
\begin{equation}
    D(C_4) \mathcal{H}(\bs{R}) D(C_4)^{\dagger} = \mathcal{H}(-\bs{R}).
\end{equation}
Given the form of a Hamiltonian, these commutation relations fix the possible $\bs{R}$-dependence of the matrix elements of the Hamiltonian.

\section{Numerical Computation of the Euler Class}
\label{app:euler_comp}
Computation of the Euler Class for a 3-level system is theoretically straightforward, but practical implementation is subtle. 
We here briefly outline the necessary steps to calculate the Euler Class numerically.

The first step is to define the grid of points which represent the $R_1, R_2$ coordinate plane upon which the Hamiltonian depends. In doing so the size of the mesh is dictated, and it is necessary to ensure this resolution is sufficient for accurate calculations. All results in the main text were obtained using a 400$\times$400 point mesh, although sufficient accuracy to obtain qualitatively correct results requires as little as a 60$\times$60 point mesh. For the sake of generality, we will take $R_1$ to be indexed by $n\in[0,n_{max}]$ and $R_2$ by $m\in[0,m_{max}]$ for the following discussion. 
To avoid numerical instabilities in case a mesh point coincides with a point of degeneracy, an infinitesimal shift is introduced to the mesh. This shift is on the order of $10^{-6}$ and is added to both the $R_1$ and $R_2$ element of each coordinate in the mesh. 

Next we obtain the energies and eigenstates of each coordinate by computing the Hamiltonian and performing numerical diagonalization. We adhere to the convention that for each point the eigenvalues are ordered by ascending energy, with their corresponding eigenvectors likewise ordered. It is necessary to perform gauge smoothing on the lowest two eigenstates, which are the states of interest in our model. The initial lack of a smooth gauge  arises from the fact that numerical calculation of eigenvectors  results in an arbitrary choice of phase for eigenvectors at each point in the mesh. Because our eigenvectors are real, this manifests as multiplication of each eigenvector at each point by a random factor of $\pm1$. While the Euler Class itself is gauge invariant, a consistent choice of gauge must be enforced to ensure its accurate computation, as  $\AA_{12}$ is not gauge invariant. The basic procedure in Python which follows must be performed individually for each state of interest, in this case state 1 and 2.
\begin{figure*}[t]
\includegraphics[width=0.95\columnwidth]{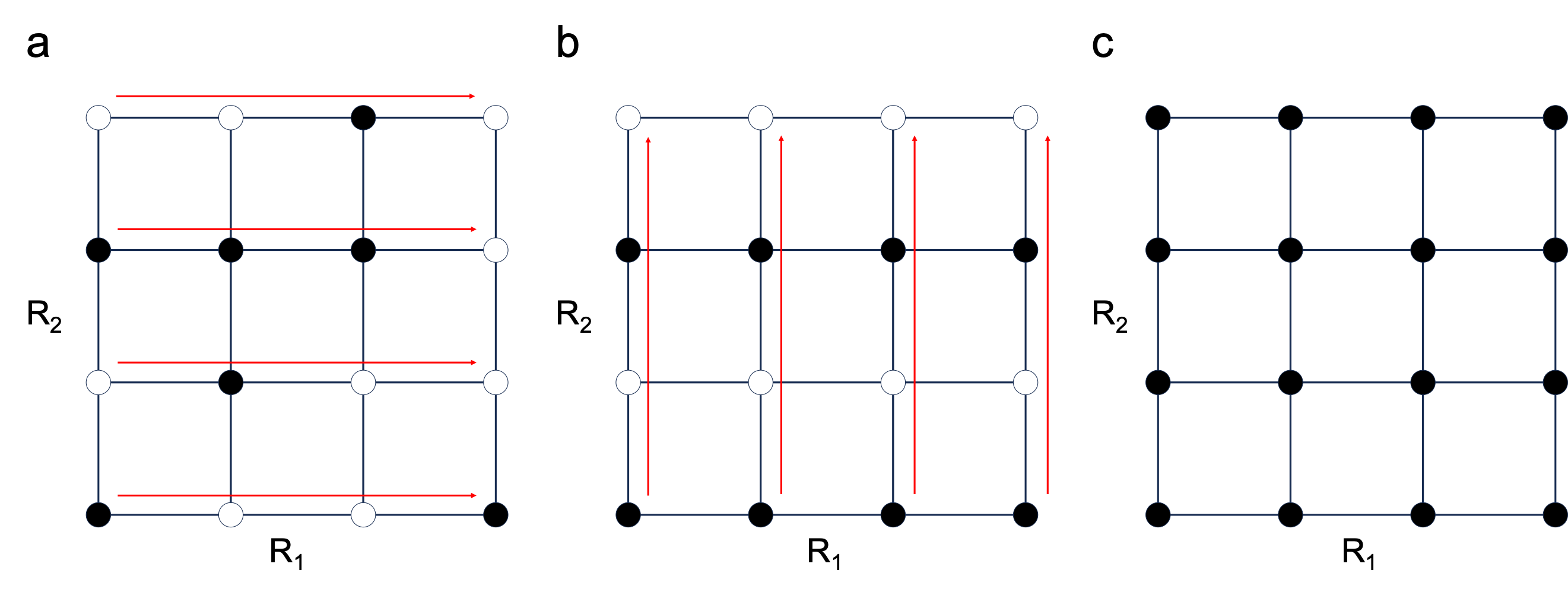}
\caption{4$\times$4 grids representing a section of the $R_1\times R_2$ mesh. The filling of each grid point represents the random phase that will be numerically assigned to an eigenstate $\ket{\psi_n}$ at that point. Filled points represent a phase of $+1$ while unfilled points represent a phase of $-1$. (a) Initially, the phase of the eigenstate at each point is random. The red arrow represents the process of matching phases along $R_1$. (b) After performing smoothing along $R_1$, there is still a mismatch in phase along $R_2$. Red arrows representing the process of matching phases along $R_2$. (c) The result of sequential gauge smoothing along $R_1$ then $R_2$ is that the sign of $\ket{\psi_n}$ is consistent for any coordinate.}
\label{fig:smooth}
\end{figure*}
Starting from the boundary of the mesh $(m,0)$ for each $m$ we iterate over all $n<n_{max}$ and enforce the condition 
\begin{equation}
    \label{eq:19}
    \braket{\psi(m,n)|\psi(m,n+1)} > 0
\end{equation}
Generally multiplying the ket by -1 if the condition is not met. This achieves smoothing along $R_2$ only, and can be considered the 'horizontal' step of the procedure as illustrated in Fig.~\ref{fig:smooth} (a)-(b). Then an accompanying 'vertical' step must be performed where starting from mesh boundary $(0,n)$ for each $n$ we iterate over all $m<m_{max}$ and enforce the condition
\begin{equation}
    \label{eq:20}
    \braket{\psi(m,n)|\psi(m+1,n)} > 0
\end{equation}
This step is illustrated in Fig.~\ref{fig:smooth} (b)-(c), and results in a smooth gauge. 
If a pair of CIs are present in the mesh, a Dirac String must be defined along a path connecting the two CIs. This constitutes assigning an opposite phase to eigenstates corresponding to each mesh point along this path after the gauge has been previously smoothed. There are two distinct steps needed to implement this change. The first is location of the CIs. It is known that transporting a wavefunction in a closed path which circumnavigates an energy level crossing results in acquisition of a $\pi$ phase. Thus computation of the Berry phase of each unit square in the grid yields a value of -1 for each square containing a CI, while a value of 1 obtains for all other squares. Given a grid of sufficiently high resolution, this phase $e^{i\gamma}$ for a state $\ket{\psi}$ around an infinitesimal square can be approximated by
\begin{equation}
e^{i\gamma} = \bra{\psi(m,n)}P(m+1,n) \cdot P(m+1,n+1) \cdot  P(m,n+1) \cdot P(m,n)\ket{\psi(m,n)}, 
\end{equation}
where $P(m,n) = \ket{\psi(m,n)}\bra{\psi(m,n)}$ is the projector on $\ket{\psi(m,n)}$. Due to the small shift introduced to the grid, we have insured that no degeneracies will lie directly on any such square, and additionally if a positive value is used for the shift to both $R_1$ and $R_2$, then the 'true' location of the degeneracy will be closest in proximity to the upper right point of the grid square containing it. This has been observed to be numerically relevant in later calculations, as the termini of the Dirac String should be as close to the actual point of degeneracy as possible. Once the coordinates for the CIs are found, defining the string is a simple matter of selecting an arbitrary path between them and altering the phase of the wavefunctions along this path.  With the system thus conditioned calculation of the Euler Class via practical implementation of Eq.~\ref{eq:euler_def} may proceed.
This calculation is broken into two steps, one for each of the integrals in Eq.~\ref{eq:euler_def}. First, we consider calculation of the integral over the Euler Connection. The path of this integral is arbitrary  so long as it encloses our region of interest and all topologically relevant features. Thus we choose it to be the rectangular boundary of our mesh. This also has the desirable consequence of ensuring the vector potential $\AA_{12}$ has only a single non-zero component along each edge of the boundary. This last feature allows the integration to be practically conducted as if over a scalar potential:

\begin{equation}
    \sum_{i}\int_{\ell_i}\AA_{12}(\bs{R})\cdot d\bs{R} = \sum_{i}\int_{\ell_i}A^j_{12}(\bs{R})dR_j , 
\end{equation}
Where $\ell_i$ represents the i$^th$ segment of the boundary, and $A^j_{12}$ represents the single nonzero component of $\AA_{12}$ over said segment. These components are calculated directly for each grid point by a simple first order forward difference approximation, which is found to yield adequate numerical results. Subsequent summation yields the Euler Connection over the boundary.
We also require the integral of the Euler Form over the region, for which we follow the methodology outlined by~\citet{bouhon2020non}.
First, a complexified state must be formed from the two states of interest as
\begin{equation}
    \ket{\psi_{12}(\bs{R})} = \frac{1}{\sqrt{2}}(\ket{\psi_1(\bs{R})}+i\ket{\psi_2(\bs{R})}),
\end{equation}
which allows definition of a complexified projector
\begin{equation}
    P_{12}(\bs{R}) = \ket{\psi_{12}(\bs{R})}\bra{\psi_{12}(\bs{R})}.
\end{equation}
Computing the Euler Form for a unit square on the mesh is performed using these projectors as
\begin{equation}
    \label{eq:25}
    \mathrm{Eu}_{ij}(m,n) =  \bra{\psi_{12}(m,n)}P_{12}(m+1,n) \cdot P_{12}(m+1,n+1) \cdot  P_{12}(m,n+1) \cdot P_{12}(m,n)\ket{\psi_{12}(m,n)}
\end{equation}
Performing this for all unit squares in the mesh, that is for $m<m_{max},n<n_{max}$ and summing yields the Euler Form. There is a single difficulty in this step, namely that the Dirac String in the real system should be of infinitesimal width, and should thus not contribute to the Euler Form. With a finite mesh this is not the case, as any unit square on the grid that possessed an edge along the Dirac string will be reversed in sign. Therefore one must ensure that Eq.~\ref{eq:19} and Eq.~\ref{eq:20} are satisfied (with respect to the complexified states) for each step of Eq.~\ref{eq:25}, multiplying by an overall sign of $-1$ whenever these conditions are violated. With the Euler Form thus calculated, the Euler Class is obtained by summing with the Euler Connection.

\end{document}